%% file: Asilomar_main.tex
\documentclass[conference]{IEEEtran}
\IEEEoverridecommandlockouts

\usepackage{cite}
\usepackage{amsmath,amssymb,amsfonts}
\usepackage{algorithm, algorithmic}
\usepackage{graphicx}
\usepackage{textcomp}
\usepackage{xcolor}
\usepackage{authblk}
\usepackage{url}
\usepackage{bbm}
\usepackage{array}

\def\BibTeX{{\rm B\kern-.05em{\sc i\kern-.025em b}\kern-.08em
    T\kern-.1667em\lower.7ex\hbox{E}\kern-.125emX}}

\input{definition}

\newcommand\numberthis{\addtocounter{equation}{1}\tag{\theequation}}
\newcommand{\norm}[1]{\left\lVert#1\right\rVert}

\begin{document}

\title{ADMM for Downlink Beamforming in Cell-Free Massive MIMO Systems
\thanks{This work is supported by the National Science Foundation under Grants \#2211815, \#2213568, \#2120447,  and the Google Scholar Research Award.}
}
\input{authors}

\maketitle

\input{abstract}

\input{V2_sections/sec_intro}
\input{V2_sections/sec_system_model}

\input{V2_sections/sec_dist_opt}

\input{V2_sections/sec_results}

\input{V2_sections/sec_conclusion}

\bibliographystyle{unsrt}
\bibliography{ref}

\end{document}

%% file: definition.tex
\def\mrm{\mathrm}
\def\mbf{\mathbf}
\def\beq{\begin{equation}}
\def \eeq{\end{equation}}
\def\bbmat{\begin{bmatrix}}
\def\ebmat{\end{bmatrix}}
\def\boldsym{\boldsymbol}

\def\what{\widehat}

\def\R{{\mathbb{R}}}
\def\C{{\mathbb{C}}}

\def\cn{\mathcal{CN}}

%% file: authors.tex

\author[1]{Mehdi Zafari}
\author[1]{Divyanshu Pandey}
\author[1]{Rahman Doost-Mohammady}
\author[1,2]{César A. Uribe}
\affil[1]{Department of Electrical and Computer Engineering, Rice University}
\affil[2]{The Ken Kennedy Institute, Rice University}


%% file: abstract.tex

\begin{abstract}

In cell-free massive MIMO systems with multiple distributed access points (APs) serving multiple users over the same time-frequency resources, downlink beamforming is done through spatial precoding. Precoding vectors can be optimally designed to use the minimum downlink transmit power while satisfying a quality-of-service requirement for each user. However, existing centralized solutions to beamforming optimization pose challenges such as high communication overhead and processing delay. On the other hand, distributed approaches either require data exchange over the network that scales with the number of antennas or solve the problem for cellular systems where every user is served by only one AP. In this paper, we formulate a multi-user beamforming optimization problem to minimize the total transmit power subject to per-user SINR requirements and propose a distributed optimization algorithm based on the alternating direction method of multipliers (ADMM) to solve it. In our method, every AP solves an iterative optimization problem using its local channel state information. APs only need to share a real-valued vector of interference terms with the size of the number of users. Through simulation results, we demonstrate that our proposed algorithm solves the optimization problem within tens of ADMM iterations and can effectively satisfy per-user SINR constraints.

\end{abstract}

\begin{IEEEkeywords}
ADMM, Distributed Optimization, Downlink Beamforming, Cell-free massive MIMO
\end{IEEEkeywords}

%% file: V2_sections/sec_intro.tex

\section{Introduction}

Cell-free massive MIMO systems seek network densification, where multiple distributed access points (APs), each equipped with a large number of antennas and connected to a central unit, collaboratively serve a group of user equipment (UEs) in a wide geographic area without any notion of cell boundaries \cite{zheng2024mobile,ngo2017cell}.
Traditionally, cell-free networks use a fully connected wireless architecture with centralized processing, control, and storage of data~\cite{dist_BF_for_cell-free}.
Such centralized network operations mitigate the adverse effects of non-coordinated collisions and interference among transmitted signals.
Moreover, in such architecture, fast fronthaul/backhaul links connect all APs to an edge cloud processor that is responsible for downlink (uplink) beamforming design for transmitting (receive) signals to (from) different UEs~\cite{rajapaksha2021deep}.
However, fully centralized cell-free networks suffer from high computational complexity and processing delays, especially when the number of antennas at each AP and the number of UEs utilizing the same time-frequency resources increase (i.e., scalability issues).
In this regard, employing distributed signal processing, optimization, and learning techniques has garnered much attention to achieve a low-latency, energy-efficient performance in a cell-free wireless network~\cite{gouda2024combined, 10159406}.
In particular, using distributed approaches can significantly reduce the complexity of solving beamforming problems (e.g., obtaining beamforming vectors at the central processor) while also significantly lowering the communication overhead.

Distributed beamforming in massive MIMO systems is done through spatial precoding, where beamformers are designed based on user channels~\cite{sam2022}.
In existing approaches, complete channel state information (CSI) should be reported to a central server by all cooperating APs to find the optimal precoder.
However, with an increasing number of antennas at each AP, complete CSI often takes the form of a large complex matrix, which is cumbersome to communicate.
Thus, ideally, APs should be enabled to locally determine their precoder with minimal information exchange with the central server and other APs.
This paper concentrates on delivering such a distributed solution.

We consider the precoder design problem with the objective of minimizing the total transmit power by all APs while ensuring a minimum signal-to-interference-noise ratio (SINR) guarantee to each UE.
Our proposed approach utilizes the alternating direction method of multipliers (ADMM) to break the precoder optimization problem into smaller optimization problems solved by each AP locally, requiring less information exchange with the central server to reach convergence.
The crux of the idea is that any given AP does not need to know the entire CSI matrix, only the total multi-user interference that other APs can cause to a UE that it intends to serve.
To achieve this, the only information that each AP needs to share with the server is the total interference it causes all UEs to experience.
When received from all APs, the central server can process this limited information and share with each AP only the cumulative interference caused by all other APs.
The APs can locally use it to find an optimal precoder.
Thus, if each AP has $N$ antennas and serves $K$ users, rather than sharing an $N \times K$ complex channel matrix with the server as in the central case, in our approach, the APs share only a real vector of size $K$, thereby significantly reducing the communication overhead.

In~\cite{ADMM}, the authors solve the downlink beamforming problem using the ADMM technique in a multiple input single output (MISO) multi-cell network, where base stations (BSs) only exchange interference values instead of complete CSI. 
ADMM was also used in~\cite{joshi2012misoadmm} to minimize the total transmit power subject to SINR per user for a MISO system. Authors in~\cite{shen2012robustadmm} provided a robust ADMM approach for coordinated beamforming in cellular systems, where imperfect CSI is considered. 
However, the network architecture considered in~\cite{ADMM, joshi2012misoadmm, shen2012robustadmm} is a multi-cell cellular network where one BS serves each user at a time.
Thus, the problem does not deal with the superposition of downlink signals from different BSs, as in our formulation.
We use the same intuitions as in~\cite{ADMM, joshi2012misoadmm, shen2012robustadmm} and extend the ADMM-based solution to a general cell-free system.
ADMM-based methods have also been used in multi-group multicast beamforming~\cite{admm_multicast_bf}, as well as recently in reconfigurable intelligent surface (RIS) aided cell-free MIMO systems~\cite{zhang2023joint, qiao2024novel} for passive beamforming design by optimizing RIS elements.

Further, in~\cite{dist_DNN_zaher}, the downlink power allocation problem is considered for a cell-free massive MIMO system.
The optimization problem considered in this work maximizes the achievable spectral efficiency (SE), 
and by solving that, training data for a deep neural network (DNN) is generated.
In~\cite{dist_PA_maxmin}, the authors consider the optimization problem for the max-min fairness of the achievable SE in a cell-free massive MIMO network.
They employ centralized and distributed DNNs to find the optimal power allocation coefficients.
\cite{dec_bf_unsup_learning} considers the optimal beamforming problem in a massive MIMO system without cells.
The authors proposed two unsupervised DNN architectures, fully and partially distributed, that can perform decentralized coordinated beamforming with zero or limited communication overhead between APs and network controllers. 
However, the challenge with employing DNN or any supervised learning method to solve optimization problems is acquiring comprehensive training data, which can be prohibitive in large-scale communication systems.
In contrast, our algorithm provides a direct iterative solution to the optimization problem without training a model.


%% file: V2_sections/sec_system_model.tex

\section{System Model and Problem Formulation}

\begin{figure}[!t]
    \centering
    \includegraphics[width=2.2in]{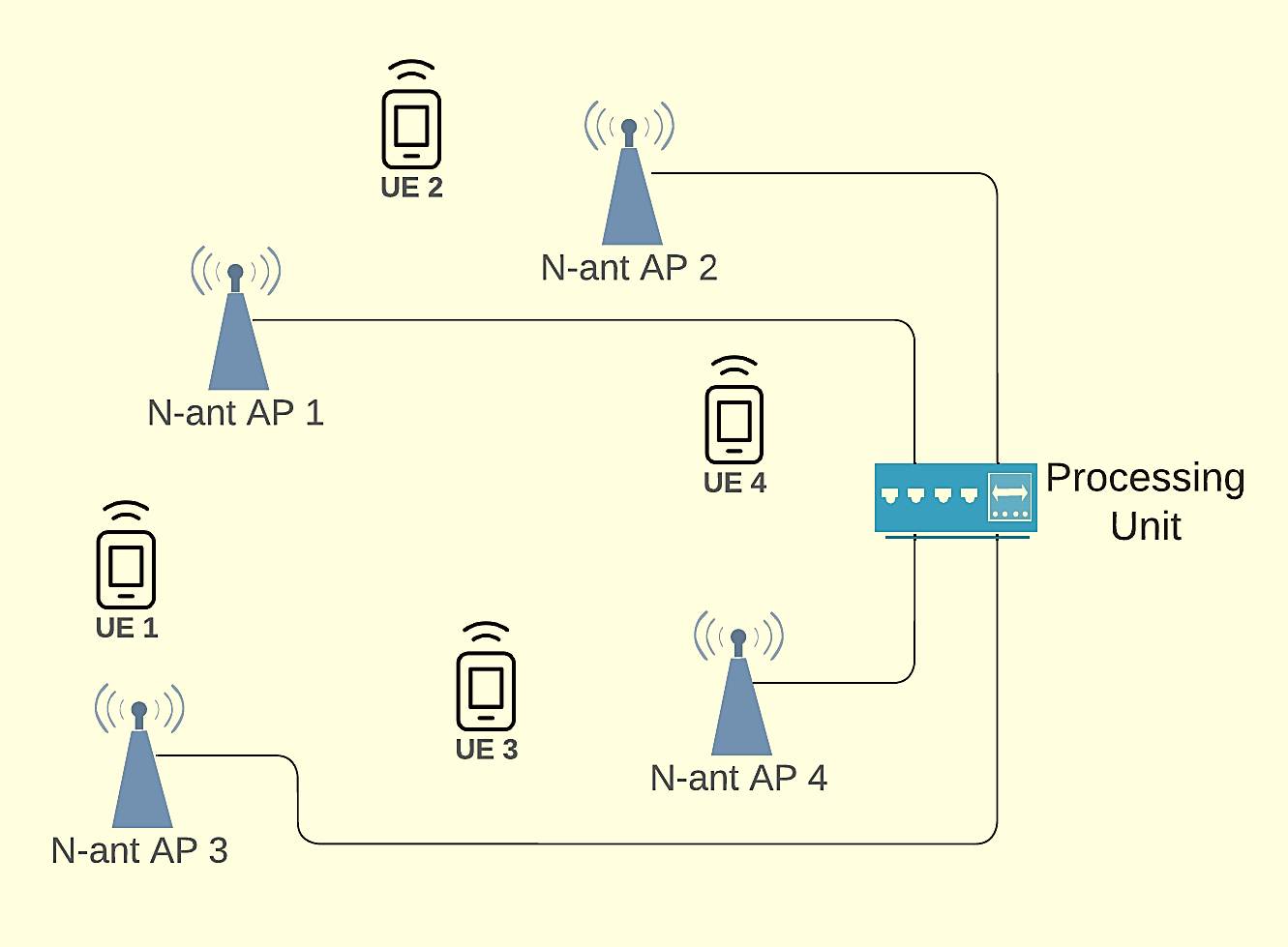}
    \caption{Illustration of the reference network architecture, with distributed APs connected to a central server and jointly serving many users.}
    \label{fig:net_arch}
\end{figure}

We consider a system consisting of \(M\) APs, each equipped with \(N\) antennas and \(K\) single-antenna users distributed over a given coverage area.
We assume that all APs are connected to a central server with enough computational capabilities via a wired fronthaul network. All \(M\) APs collaborate jointly and cooperatively to serve all \(K\) users in the joint coverage area over the same time-frequency resource.
The transmission operation follows the time division duplex (TDD) standard, where pilot, uplink data, and downlink data transmissions are separated in the time domain.
Figure~\ref{fig:net_arch} illustrates the general architecture of the reference cell-free network considered in this work.
The connection between APs follows the star topology, where all APs have a fast, error-free wired connection to the central node (processing unit).
There is no direct wired connection between the APs together.
However, they can exchange information over the air or through the central node.

The channel from user \(k\) to antenna \(i\) at AP \(m\) over each OFDM subcarrier with the width of \(\Delta f = \mrm{BW}/N_{sc}\) is considered a narrowband slow-fading channel denoted with the complex gain \(\mrm{h}_{km}^{(i)} \in \C\). In a propagation environment with non-line-of-sight (NLOS) channels, the distribution of the channel gain \(\mrm{h}_{km}^{(i)}\)can be considered as zero-mean circular-symmetric complex Gaussian, which results in a Rayleigh distribution for the magnitude and Uniform distribution for the phase of the channel gain.
The channel vector between user \(k\) and AP \(m\) is denoted as
$
    \mrm{\mbf{h}}_{km} = [\mrm{h}_{km}^{(1)}\, \mrm{h}_{km}^{(2)}\, \cdots \, \mrm{h}_{km}^{(N)}]^\top
\in \mathbb{C}^N$.
Channel statistics depend on characteristics of the propagation environment, including large-scale fading coefficient, shadowing, and other spatial properties.
In our formulation, we assume that each AP locally knows the channel estimates through uplink pilot training.
Precoding vectors can be designed centralized, with all APs required to send their local channel estimates to the central server for computations.
An alternate approach is distributed precoder design, where APs need not share their complete local channel estimates with the central server but may exchange only a limited amount of information.

In the centralized approach, the local CSI estimated by each AP is sent to the central server to calculate the optimal precoding vectors for each AP.
Let the precoding vector for the transmission from AP \(m\) to user \(k\) be denoted by \(\mathrm{\mathbf{w}}_{km} \in \mathbb{C}^{N}\).
The received downlink signal at user \(k\) can be expressed as
\begin{align*}
    y_k^{\mathrm{dl}} &= \sum\limits_{m=1}^{M} \mathrm{\mathbf{h}}_{km}^\top\, \mathrm{\mathbf{x}}_{m} + \mathrm{n}_k 
    = \sum\limits_{m=1}^{M} \, \sum\limits_{u=1}^{K} \mathrm{\mathbf{h}}_{km}^\top \, \mathrm{\mathbf{w}}_{um} \, \mrm{s}_u + \mathrm{n}_k.
    \numberthis
\end{align*}
The vector \(\mathrm{\mathbf{x}}_{m} \in \mathbb{C}^{N} \) is the precoded signal transmitted by AP \(m\),
\(\mrm{s}_u \in \mathbb{C}\) is the data signal for user \(u\), and \(\mathrm{n}_k\) is the additive noise signal at user $k$, modeled as a circular symmetric complex Gaussian with zero mean and variance \(\sigma_k^2\). 
Without loss of generality, we assume that the input signal is normalized to unit power such that \(\mathbb{E}[|\mrm{s}_k|^2] = 1, \; \forall k\).
Thus, for a given channel vector and precoder, we can write the instantaneous SINR at the $k$th user as
\begin{equation}
    \mathrm{SINR}_k = \frac{|\sum_{m=1}^{M} \mathrm{\mathbf{h}}_{km}^\top \, \mathrm{\mathbf{w}}_{km}|^2}{\sum_{{u=1},\, {u\neq k}}^{K} |\sum_{m=1}^{M} \mathrm{\mathbf{h}}_{km}^\top \, \mathrm{\mathbf{w}}_{um}|^2 + \sigma_k^2}.
\end{equation}
Since the input signals are normalized to unit power, the vectors $\mathrm{\mathbf{w}}_{km}$ essentially represent both operations, precoding and power allocation.
Hence, the total transmit power from $m$th AP can be written as $ \sum_{k=1}^{K} \norm{\mathrm{\mathbf{w}}_{km}}^2$. Subsequently, the total downlink transmit power from all APs can be given as $\sum_{m=1}^{M} \sum_{k=1}^{K} \norm{\mathrm{\mathbf{w}}_{km}}^2$.

Our objective is to design the downlink precoding vectors $\mathrm{\mathbf{w}}_{km}$ such that the total transmitted power from all APs is minimized while satisfying some quality of service (QoS) constraint for all users.
The QoS considered in this work is in terms of minimum SINR guarantees for each user.
Since the rate is a monotonically increasing function of SINR, the minimum SINR requirement is equivalent to the minimum rate required by each user.
Thus, the objective can be stated as the following optimization problem:
\begin{subequations}\label{eq:opt_prob}
\begin{alignat*}{2}
    &\underset{\{\mathrm{\mathbf{w}}_{km}\}}{\mathrm{minimize}} \quad \sum\limits_{m=1}^{M} \sum\limits_{k=1}^{K} \norm{\mathrm{\mathbf{w}}_{km}}^2 \numberthis \\
    &\mathrm{s. t.} \;  \frac{|\sum_{m=1}^{M} \mathrm{\mathbf{h}}_{km}^{\top} \mathrm{\mathbf{w}}_{km}|^2}{\sum\limits_{{u=1}, {u\neq k}}^{K} |\sum_{m=1}^{M} \mathrm{\mathbf{h}}_{km}^{\top} \mathrm{\mathbf{w}}_{um}|^2 + \sigma_k^2} \geq \gamma_k, \, \forall k \numberthis \label{eq:opt_prob_const}
\end{alignat*}
\end{subequations}
where \(\gamma_k\) is the minimum required SINR at user \(k\), and the inequality constraint should be satisfied for all \(k=1, ..., K\).
The centralized QoS-constrained power minimization problem~\eqref{eq:opt_prob} has a strongly convex objective function.
The constraint on SINR is generally non-convex because it is the ratio of two quadratic functions.
However, it can be transformed or approximated into a convex form under certain conditions.
In the subsequent section, we approximate the SINR constraint \eqref{eq:opt_prob_const} and reframe the centralized problem \eqref{eq:opt_prob} so that ADMM can be applied to solve it in a distributed manner.

%% file: V2_sections/sec_dist_opt.tex

\section{Distributed Solution for Precoder Optimization}

Solving \eqref{eq:opt_prob} in a centralized setting requires all APs to send their downlink estimated channel vectors (local CSI) to the central server.
Since the objective function in \eqref{eq:opt_prob} is convex, by having channel vectors at the server, different convex optimization approaches can be employed to find the optimal solution.
However, as discussed earlier, exchanging the channel estimates and the precoding vectors between APs and the central server poses several challenges, especially with a large number of users or APs.
The computational complexity of finding precoding vectors at the central server and the overhead of information exchange between APs and the server scale drastically with increased network size.
Alternatively, employing a distributed solution can reduce communication overhead on the network and allow concurrent utilization of computational resources at each AP to find the solution.
Furthermore, an additional benefit of a distributed solution is its potential to continuously adapt to small changes in CSI with minimal information exchange once it has converged~\cite{ADMM}.

To apply a distributed optimization technique, it is necessary to reformulate~\eqref{eq:opt_prob} as a decomposable optimization problem.
The objective function can be written as
\begin{equation}
    \underset{\{\mathrm{\mathbf{W}}_m\}_{m=1}^{M}}{\mathrm{minimize}} \quad \sum_{m=1}^{M}\, f_m\left(\mathrm{\mathbf{W}}_m\right),
\end{equation}
where the matrix \(\mathrm{\mathbf{W}}_m = \left[\mathrm{\mathbf{w}}_{1m}, ..., \mathrm{\mathbf{w}}_{Km}\right] \in \mathbb{C}^{N\times K}\) contains the precoding vectors from the $m$th AP to all \(K\) users.
The function $f_m( \mathbf{W}_m)$ denotes the total transmit power by the $m$th AP and is given as
\begin{equation}
f_m\left(\mathrm{\mathbf{W}}_m\right) = \sum_{k=1}^{K} \norm{\mathrm{\mathbf{w}}_{km}}^2 = \mathrm{tr} \left(\mathrm{\mathbf{W}}_m^{\mathrm{H}} \mathrm{\mathbf{W}}_m\right),
\end{equation}
which can be computed by the $m$th AP independent of other APs.
However, the inequality constraints \eqref{eq:opt_prob_const} on the instantaneous SINR of each user cannot be separated in the AP index \(m\) since each AP needs the interference information from other APs to measure the SINR for any user. 
Therefore, in order to distribute the optimization problem \eqref{eq:opt_prob} over APs, the inequality constraint must be modified so that it can be decomposed into \(m\), allowing the use of the alternating direction method of multipliers (ADMM) to solve the optimization problem iteratively in a distributed manner.

The product \(\mathbf{h}_{km}^\top\mathbf{w}_{um} \in \mathbb{C}\), which represents the interference that the signal intended for user $u$ by AP $m$ will cause to user $k$, is a complex scalar and is available for all \(u=1,...,K\) at only AP \(m\).
To modify the SINR constraint \eqref{eq:opt_prob_const}, we first take the square root of it, $\sqrt{\mathrm{SINR}_k}$, as
\begin{equation}
    \label{eq:constraint_sqrt}
    \frac{|\sum_{m=1}^{M} \mathrm{\mathbf{h}}_{km}^\top \mathrm{\mathbf{w}}_{km}|}{\left(\sum_{{u=1}, {u\neq k}}^{K} |\sum_{m=1}^{M} \mathrm{\mathbf{h}}_{km}^\top \mathrm{\mathbf{w}}_{um}|^2 + \sigma_k^2\right)^{1/2}} \geq \sqrt{\gamma_k}.
\end{equation}
Following the triangle inequality, the numerator in \eqref{eq:constraint_sqrt} can be bounded as
\begin{equation}\label{eq:traingle_ineq}
    \left|\sum\nolimits_{m=1}^{M} \mathrm{\mathbf{h}}_{km}^\top \mathrm{\mathbf{w}}_{km}\right| \leq \sum\nolimits_{m=1}^{M} \left|\mathrm{\mathbf{h}}_{km}^\top \mathrm{\mathbf{w}}_{km}\right|.
\end{equation}
The left-hand side of \eqref{eq:traingle_ineq} denotes the total magnitude gain of the desired signal at a given user \(k\), which is a superposition of the transmitted signals from all APs and is upper bounded by the sum of the individual gains of the desired signal from all APs.
A key feature of massive MIMO systems is favorable propagation, which implies that channels from APs to users tend to be nearly orthogonal.
Assuming that all the APs perform massive MIMO operations and there is sufficient spatial separation between them, we consider the channel vectors from each user to different APs to be uncorrelated.
Thus, since the precoding vectors are calculated based on the channel vectors, the phase of the complex scalar \(\mathbf{h}_{km}^\top\mathbf{w}_{um}\) for different \(m\) tends to be equal, making the difference between two sides of the triangle inequality very small. This intuition leads us to relax the constraint by considering the upper bound of the desired signal.
Subsequently, we argue that a feasible SINR boundary exists, using which satisfying the relaxed constraint results in satisfying the actual SINR constraint.

The denominator in \eqref{eq:constraint_sqrt} can be bounded using the generalized form of Minkowski's inequality \cite{marshall2011inequalities} as
\begin{multline}
    \label{eq:constraint_den}
    \left(\sum\nolimits_{{u=1}, {u\neq k}}^{K} \left|\sum\nolimits_{m=1}^{M} \mathrm{\mathbf{h}}_{km}^\top \mathrm{\mathbf{w}}_{um}\right|^2 + \sigma_k^2\right)^{1/2} \\
    \leq \sum\limits_{m=1}^M \left(\sum\nolimits_{{u=1}, {u\neq k}}^{K}\left|\mathbf{h}_{km}^\top\mathbf{w}_{um}\right|^2\right)^{1/2} + \sigma_k,
    \numberthis
\end{multline}
where the upper bound can be separated in \(m\).
Transmitted signals from all APs cause inter-user interference at each user.
However, at each given AP \(m\), the knowledge of the channel and precoding vectors of other APs is unavailable.
To address this issue, we introduce an auxiliary variable \(\mathrm{I}_{km}\) as the total inter-user interference caused by AP \(m\) at user \(k\).
Using this auxiliary variable, we can rewrite the upper bound in \eqref{eq:constraint_den} for a given AP \(m\) as
\begin{equation}
    \left(\sum\limits_{{u=1}, {u\neq k}}^{K}\left|\mathbf{h}_{km}^\top\mathbf{w}_{um}\right|^2\right)^{1/2} + \sum\limits_{j=1, j\neq m}^M \mathrm{I}_{kj} + \sigma_k,
\end{equation}
where the first term is the inter-user interference caused by AP \(m\) to user $k$ and can be calculated locally by the $m$th AP.
The second term is the total interference caused by the other $M-1$ APs to user $k$.
The inter-user interference caused by $j$th AP to user $k$ is given as
\begin{equation}
    \mathrm{I}_{kj} \triangleq \left(\sum\limits_{u=1, u\neq k}^{K}\left|\mathbf{h}_{kj}^\top\mathbf{w}_{uj}\right|^2\right)^{1/2},
\end{equation}
which can be measured at AP \(j\).
Although this interference term is not available at AP \(m\) for \(m\neq j\), we introduce a copy of this term at AP \(m\) denoted by \(\mathrm{I}_{kj}^{(m)}\) and further we impose the constraint that \(\mathrm{I}_{kj}^{(m)} = \mathrm{I}_{kj}^{(j)}\), i.e., the interference copy should be equal to the measured value.
Each AP will have a copy of this interference value, and they will exchange this information through the central node until they finally reach a consensus.
Exchanging interference values has a much lower communication load on the fronthaul network than exchanging the channel vectors for all users.
Moreover, it does not scale with \(N\), the number of antennas at each AP, which is an important factor in massive MIMO systems.

To enable the use of the ADMM algorithm, we need to impose the equality of the measured interference values, i.e., \(\mathrm{I}_{km}^{(m)}\) for the interference caused by AP \(m\) measured at AP \(m\), and \(\mathrm{I}_{km}^{(j)}\) for the copy of interference caused by AP \(m\) copied at all other APs \(j\neq m\).
Using these notations, the optimization problem \eqref{eq:opt_prob} can be modified and reformulated as \eqref{eq:opt_modified}, similar to the formulation in \cite{ADMM}.
The constraint \eqref{eq:opt_modified_const1} is the relaxed version of the original SINR constraint \eqref{eq:opt_prob_const}, which is now decomposable in \(m\).
This modified constraint is obtained by replacing the denominator of the left-hand side of \eqref{eq:constraint_sqrt} with its upper bound, as given in \eqref{eq:constraint_den}, along with taking advantage of the favorable propagation property of massive MIMO systems to consider uncorrelated user channels and relax the numerator of the right-hand side of \eqref{eq:constraint_sqrt} using the inequality given in \eqref{eq:traingle_ineq}.

\begin{subequations}\label{eq:opt_modified}
\begin{alignat*}{2}
    &\underset{\{\mathrm{\mathbf{w}}_{km}\},\{\mathrm{I}_{km}\}}{\mathrm{minimize}} \quad \sum\limits_{m=1}^{M} \sum\limits_{k=1}^{K} \norm{\mathrm{\mathbf{w}}_{km}}^2 \numberthis \\
    &\mathrm{s. t.} \; \frac{\sum\limits_{m=1}^{M} \left|\mathrm{\mathbf{h}}_{km}^\top \mathrm{\mathbf{w}}_{km}\right|}{\left(\sum\limits_{u\neq k}^{K}\left|\mathbf{h}_{kj}^\top\mathbf{w}_{uj}\right|^2\right)^{1/2} + \sum\limits_{m\neq j}^M \mathrm{I}_{km}^{(j)} + \sigma_k} \geq \widehat\gamma_k, \, \forall k \numberthis \label{eq:opt_modified_const1}\\
    & \qquad \mathrm{I}_{km}^{(m)} \geq \left(\sum\limits_{u\neq k}^{K}\left|\mathbf{h}_{km}^\top\mathbf{w}_{um}\right|^2\right)^{1/2}, \quad \forall k,m \numberthis \label{eq:opt_modified_const2}\\
    & \qquad \mathrm{I}_{km}^{(j)} = \mathrm{I}_{km}^{(m)}, \quad \forall k, m, j\neq m. \numberthis \label{eq:opt_modified_const3}
\end{alignat*}
\end{subequations}
Thus, by setting \(\widehat\gamma_k \geq \sqrt{c \, \gamma_k}\) for scalar \(c \geq 1\) being controlled based on the propagation environment, the original SINR constraint in \eqref{eq:opt_prob_const} will be satisfied.
Moreover, to handle the equality constraint \eqref{eq:opt_modified_const3}, we define a global consistency variable \(\Omega_k\) to ensure that all APs agree on the total interference suffered by user \(k\).
Since for calculating the SINR for each user we only need the sum of all interference values originating from each AP, we define a new variable for the sum of all interferences as $\mrm{z}_{km} \triangleq \sum_{i=1}^{M} \mrm{I}_{ki}^{(m)}$, where \(\mrm{z}_{km}\) should eventually become equal to \(\Omega_k\) for every AP \(m\); in other words, all APs need to agree on the total interference values.
In the vector form we have \(\mbf{z}_m = [\mrm{z}_{1m}, \, \mrm{z}_{2m}, \cdots, \, \mrm{z}_{Km}]^\top \, \in \R^{K}\) and \(\boldsym\Omega = [\Omega_1, \, \cdots, \, \Omega_K] \in \R^{K}\).
These variables facilitate the management of the equality constraint imposed on the sum of the interference values.
In order to write the formulation of the local problems solved by each of the APs, we introduce the function \(\mathrm{g}_{km}\) as
\begin{align*}
    &\mathrm{g}_{km} (\mathbf{H}_m, \mathbf{W}_m, \mbf{z}_m) \triangleq \\ &\frac{\left|\mathrm{\mathbf{h}}_{km}^\top \mathrm{\mathbf{w}}_{km}\right|}{\left(\sum\nolimits_{u\neq k}^{K}\left|\mathbf{h}_{km}^\top\mathbf{w}_{um}\right|^2\right)^{1/2} + \sum\limits_{i\neq m}^M \mathrm{I}_{ki}^{(m)} + \sigma_k}, \,\,\, \forall \, k,m
    \numberthis
\end{align*}
for \(\mathbf{H}_m = [\mathbf{h}_{1m}, ..., \mathbf{h}_{Km}] \in \mathbb{C}^{N\times K}\), \(\mathbf{W}_m \in \mathbb{C}^{N\times K}\), and \(\mbf{z}_m\, \in \R^{K}\) containing the channel vectors, precoding vectors, and summation of local interference values available at AP \(m\), respectively.
For AP \(m\) and user \(k\), we define
\begin{equation}
\mathrm{INT}_{km} (\mathbf{H}_m,\! \mathbf{W}_m,\! \mbf{z}_m) \triangleq \mathrm{I}_{km}^{(m)}\!-\! \left(\sum\nolimits_{u\neq k}^{K}\left|\mathbf{h}_{km}^\top\mathbf{w}_{um}\right|^2\right)^{1/2}
\end{equation}
Using these definitions, the optimization problem \eqref{eq:opt_modified} can be written in a compact form as
\begin{subequations}\label{eq:opt_modified_compact}
\begin{alignat*}{2}
&\underset{\mathbf{W}_m,\mbf{z}_m,\boldsymbol\Omega}{\mathrm{minimize}} \quad &&\sum\limits_{m=1}^{M} \mathrm{tr} \left(\mathrm{\mathbf{W}}_m^{\mathrm{H}} \mathrm{\mathbf{W}}_m\right) \numberthis \\
    & \; \mathrm{subject\; to} \; &&\sum\limits_{m=1}^M \mathrm{g}_{km} (\mathbf{H}_m, \mathbf{W}_m, \mbf{z}_m) \geq \widehat\gamma_k, \quad \forall k \numberthis \\
    & &&\mathrm{INT}_{km} (\mathbf{H}_m, \mathbf{W}_m, \mbf{z}_m) \geq 0, \quad \forall k, m \numberthis \\
    & &&\mbf{z}_m = \boldsymbol\Omega, \quad \forall m. \numberthis 
\end{alignat*}
\end{subequations}

\begin{algorithm}
\caption{Distributed Beamforming with ADMM}\label{alg:alg_admm}
\begin{algorithmic}[1]
\STATE Initialize \(\boldsymbol\Omega^{[0]}\) and \(\mathbf{\mathcal{V}}^{[0]}\). Set \(t=1\).
\STATE Distributively solve
\begin{alignat*}{2}
    &\underset{\mbf{z}_m,\mathbf{W}_m}{\mathrm{minimize}} &&\mathrm{tr} \left(\mathrm{\mathbf{W}}_m^{\mathrm{H}} \mathrm{\mathbf{W}}_m\right) + \left(\mathbf{\mathcal{V}}_m^{[t-1]}\right)^\top \left(\mbf{z}_m - \boldsymbol\Omega^{[t-1]}\right) \\
    & &&+ \frac{\rho}{2} \, \norm{\mbf{z}_m - \boldsymbol\Omega^{[t-1]}}^2 \\
    &\mathrm{subject \; to}  \quad && \mathrm{g}_{km} (\mathbf{H}_m, \mathbf{W}_m, \mbf{z}_m) \geq \frac{1}{M}\what\gamma_k, \\
    & && \mathrm{INT}_{km} (\mathbf{H}_m, \mathbf{W}_m, \mbf{z}_m) \geq 0
\end{alignat*}
to obtain \(\mathbf{w}_{km}^{[t]}\) for \(k=1,...,K\) and \(\mbf{z}_m^{[t]}\) at each AP \(m\).
\STATE  Each AP sends interference terms \(\mbf{z}_m^{[t]}\) to the central node.
\STATE The central node updates the global parameter \(\boldsymbol\Omega^{[t]}\) by taking average of interference terms received from all APs.
\STATE Update \(\mathbf{\mathcal{V}}_m^{[t]} \gets  \mathbf{\mathcal{V}}_m^{[t-1]} + \rho \left(\mbf{z}_m^{[t]} - \boldsymbol\Omega^{[t]}\right)\) at each AP \(m\).
\STATE Exit if the stopping criteria are met. \\ Otherwise, set \(t \gets t+1\) and return to 2.
\end{algorithmic}
\end{algorithm}

ADMM stands out as a notable algorithm for solving convex optimization problems in a distributed manner due to its capability to iteratively alternate between optimizing multiple variables.
Both ADMM and dual decomposition can be used to decouple problems coupled through a constraint \cite{decomposition_tutorial}.
Once the original problem \eqref{eq:opt_prob} is reformulated and decomposed via ADMM, it loses strong convexity, which is the price of making it distributed.
Assuming that the channel remains static during the time of transmission and computation, the optimization problem in \eqref{eq:opt_modified_compact} can thus be iteratively solved by Algorithm \ref{alg:alg_admm}, which represents the proposed ADMM solution for the derived downlink beamforming optimization problem.
We assume identical fronthaul communication links and computation resources at all APs, leading us to assign the same Lagrange multiplier to the residual interference terms from all other APs at every given AP \(m\).
Step 2 of Algorithm \ref{alg:alg_admm} minimizes the augmented Lagrangian for each AP $m$, where \(\mathbf{\mathcal{V}}_m\) is the vector of dual variables and \(\rho\) is the penalty parameter.
This step can be done using the standard convex optimization solvers locally at each AP.

The convergence proofs for ADMM and the framework to apply ADMM to a consensus optimization with regularization are provided in~\cite[Appendix~A]{boyd_book}.
The problem \eqref{eq:opt_modified_compact} and the solution in Algorithm~\ref{alg:alg_admm} are formulated such that the general convergence proofs apply, and the optimal solution is provided in the limit.
We omit further details due to the lack of space.
In the next section, we provide simulation results that illustrate the effectiveness of the algorithm.


%% file: V2_sections/sec_results.tex

\section{Simulation Results}

To demonstrate the performance of our algorithm, we conduct simulations on a network comprising \(M=\{2,4\}\) distributed APs, each equipped with \(N = 64\) antennas, serving \(K = 4\) single-antenna users simultaneously.
The channel model follows Rayleigh distribution, \(\mrm{\mbf{h}}_{km} \sim \cn \left(\mbf{0}, \beta_{km} \, \mbf{I}_{N}\right)\).
The large-scale fading coefficient \(\beta_{km}\) shows the average channel gain from user \(k\) to AP \(m\), and $\mbf{I}_{N}$ is an identity matrix.
The ratio of the average channel gain to the noise variance in the environment is \(\beta_{km} / \sigma_k^2 = 20\) dB for all users.
To find the optimal precoders in the centralized case, all the APs share their channel data with the central server, where the original optimization problem \eqref{eq:opt_prob} is solved.
Conversely, in the distributed setting, we leverage Algorithm \ref{alg:alg_admm} to optimize without necessitating the exchange of local CSI, relying solely on shared interference values.
To initiate the algorithm, we set the dual parameters \(\mbf{\mathcal{V}}^{[0]} = \mbf{0}\), and through empirical analysis, we determined that setting \(\rho = 10\) ensures a sufficiently fast convergence.
We run Algorithm \ref{alg:alg_admm} across over 100 channel realizations, consistently observing convergence within up to 10 ADMM iterations.

\begin{figure}[!t]
    \centering
    \includegraphics[width=3.49in]{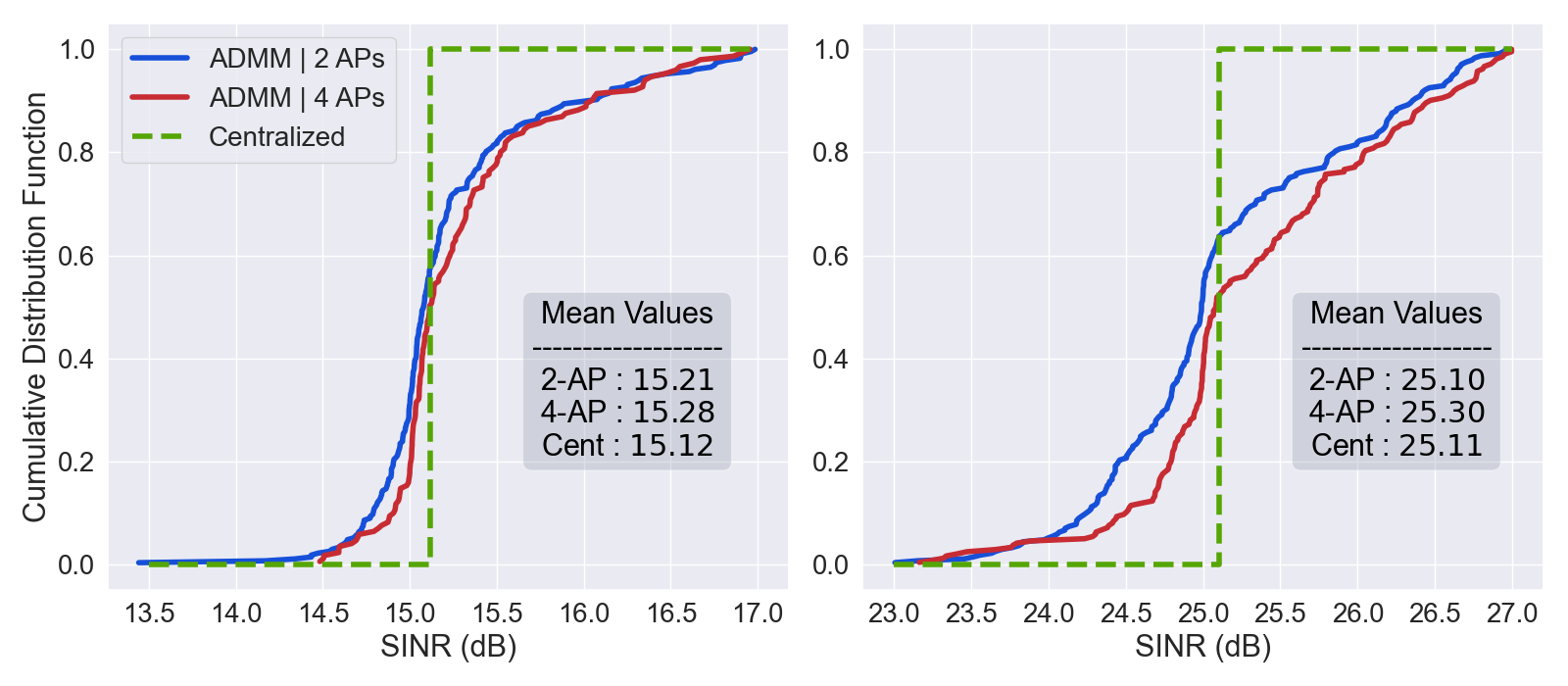}
    \caption{SINR CDF for centralized vs distributed solution (Algorithm \ref{alg:alg_admm}) under the same SINR constraint $\gamma_k$ for all of the users across channel realizations. Left: $\gamma_k=15$ dB, Right: $\gamma_k=25$ dB.}
    \label{fig:sinr_cdf}
\end{figure}

\begin{figure}[!t]
    \centering
    \includegraphics[width=3.49in]{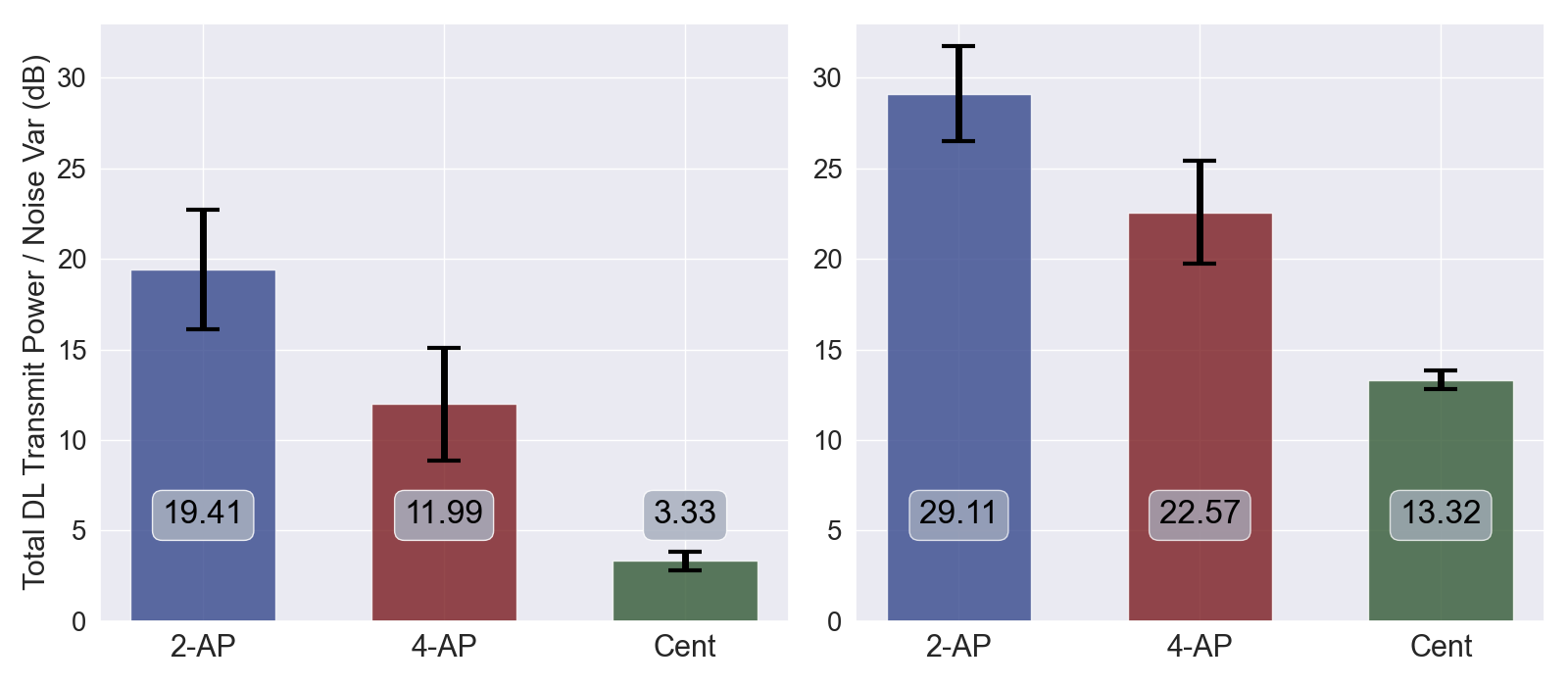}
    \caption{Total transmit SNR for distributed approach (with 2 and 4 APs), and centralized solution (with 4 APs) under same  SINR constraint $\gamma_k$ for all users. Left: $\gamma_k=15$ dB, Right: $\gamma_k=25$ dB.}
    \label{fig:power_bar}
\end{figure}

We solve the optimization problem in both centralized and distributed settings for two different SINR constraints, \(\gamma_k = \{ 15, 25 \} \) dB for each user.
For the distributed approach, we set \(\widehat\gamma_k = \sqrt{\gamma_k}\) in our algorithm.
We then measure the achieved SINR in downlink, averaged across users, using the precoders obtained as the solution.
The cumulative distribution function (CDF) of the achieved SINR for both scenarios is illustrated in Fig. \ref{fig:sinr_cdf}.
The figure on the left demonstrates that our distributed algorithm, within up to 10 ADMM iterations, achieves a precoding solution resulting in SINR higher than \(14.5\) dB for all channel realizations.
It also shows that the average SINR across channel realizations meets the constraint, with less than \(20 \%\) outage.
Note that to ensure the instantaneous SINR meets the constraint for all channel realizations such that we have no outage, we can set \(\widehat\gamma_k\) to be larger than \(\sqrt{\gamma_k}\), i.e., \(c > 1\) (in this case \(c = 1.13\) is enough).
We intentionally used the minimum value \(\widehat\gamma_k = \sqrt{\gamma_k}\) to demonstrate how much the original SINR constraint is violated due to the relaxation used.
In Fig. \ref{fig:sinr_cdf} on the right side, the performance is shown for the case of \(\gamma_k = 25\) dB minimum SINR.
In this case, the SINR provided by Algorithm \ref{alg:alg_admm} is always higher than \(23.5\) dB, and the outage is around $40\%$.
Again, to ensure no outage, we can set \(c = 1.42\).
When increasing the number of APs from 2 to 4  in both scenarios, Algorithm \ref{alg:alg_admm} still performs well and indicates even lower probability mass for SINR values less than the minimum.

Fig. \ref{fig:power_bar} demonstrates the total downlink power used in all the scenarios shown in Fig. \ref{fig:sinr_cdf}.
As expected, the minimum power is achieved by solving the optimization problem centrally at the server.
Zero-forcing or MMSE beamformers designed centrally at the server will perform no better than this optimal centralized solution.
Conversely, conjugate beamforming is a fully distributed method that can be performed locally at each AP.
Through implementing it, we achieved a total transmit SNR higher than \(40\) dB for all scenarios, where the maximum achieved SINR was \(12\) dB with 2 APs and \(14.3\) dB with 4 APs. 
Fig. \ref{fig:power_bar} also illustrates that to achieve \(10\) dB higher downlink SINR, the total downlink transmit power increases \(10\) dB correspondingly.
Algorithm \ref{alg:alg_admm} does not require CSI sharing and can achieve a performance comparable to the optimal one with a limited information exchange over the network.
Table \ref{tab:compare} provides a comparison of our proposed distributed solution to the ADMM-based algorithm in \cite{ADMM} and the centralized solutions.
As indicated in the table, the size of data to be shared is reduced by a factor of \(N\times 2\) compared to the centralized solution, where $N$ can be very large in massive MIMO systems ($N=16$ in our simulations).

\newcolumntype{M}[1]{>{\centering\arraybackslash}m{#1}}
\begin{table}[!t]
\caption{Comparing centralized vs distributed solutions. \\ ($M$ APs, $N$ Antennas per-AP, $K$ UEs)}
\label{tab:compare}
\centering
\begin{tabular}{|m{1.9cm}|M{1.4cm}|M{1.9cm}|M{1.6cm}|} 
\hline
      & \textbf{Centralized} & \textbf{\cite{ADMM}} & \textbf{Algorithm \ref{alg:alg_admm}} \\
    \hline \hline
    Network type & all & cellular & cell-free \\
    \hline
    Data to share  &  Local CSI & Magnitude of interference & Magnitude of interference  \\
    \hline
    Size of total data to be shared  &  \(\C^{M \times N \times K}\) & \(\R^{M \times (M-1) \times K}\) & \(\R^{M \times K}\)  \\
    \hline
    Processors used  &  1 & \(M\) & \(M\) \\
    \hline
    Scalable  &  NO & YES & YES \\
    \hline
\end{tabular}
\end{table}

%% file: V2_sections/sec_conclusion.tex

\section{Conclusion}

In this work, we proposed a distributed solution for downlink beamforming in a multi-user cell-free massive MIMO system with per-user SINR constraint.
We proposed an ADMM-based algorithm for the problem that requires all APs to exchange only the sum of the interference values over the network instead of complete CSI matrices, thus significantly reducing the communication overhead between the APs and the central server.
The proposed approach was analyzed through numerical simulations.
A potential future direction for this work is to analyze the algorithm's performance under imperfect channel estimates. Another direction is to explore techniques such as early termination and acceleration to speed up the convergence of the ADMM solution.